# A Decade of Healthcare Cyber Threats: Empirical Analysis, Evidence-Based Prioritisation, and AI Threat Model

Sadia Mehrin Rahi[1], Ayesha Siddika[1], Istiyak Hasan Maruf[1], Adrita Rahman Tory[1], Muhammad Aminur Rahaman[1], and Khondokar Fida Hasan[2]*

Bangladesh University of Business and Technology (BUBT), Mirpur-2, Dhaka-1216, Bangladesh
University of New South Wales (UNSW), ACT 2601, Australia
fida.hasan@unsw.edu.au

**Abstract.** Healthcare systems face persistent and evolving cyber threats, yet how adversarial tactics and techniques have shifted over time has not been systematically characterised using empirical, multi-source data. This paper analyses 1,214 threat records drawn from three authoritative sources: the MITRE ATT&CK behavioural framework, the CISA Known Exploited Vulnerabilities catalogue, and the NIST vulnerability database, covering 44 validated healthcare-targeting threat entities from 2017 to 2024. We show that attacker behaviour has shifted measurably toward stealth-oriented tactics: defense evasion remained the dominant tactic throughout the observation period, consistently accounting for 15–20% of observed technique use from 2017 to 2024, while persistence declined from 11.2% to zero and initial access from 9.0% to zero over the same period. We further demonstrate that existing detection guidance is structurally misaligned with where attacker effort is concentrated, with the least-covered techniques receiving the most adversarial attention. A convergence analysis links 679 confirmed exploited vulnerabilities to a single dominant behavioural technique, identifying a common addressable chokepoint across the vulnerability and behavioural surfaces. Finally, we identify 42 high-priority techniques representing immediate detection opportunities, and show that this technique set maps directly to emerging threats against AI-integrated clinical systems.

## 1 Introduction

Healthcare is the most targeted global critical-infrastructure sector due to three structural factors: high black-market data value, patient-safety risks restricting clinical patching, and immediate pressure to pay ransoms to avoid downtime [1]. This threat is accelerating: the HHS recorded 1,710 incidents in the year ending October 2024 (a 24,% year-on-year increase), while the FBI's 2024 Internet Crime Report ranked healthcare first among US critical infrastructure by ransomware and data-theft volume [2,3]. Concurrently, adversary behavior has shifted structurally. Financially motivated ransomware operators are now joined by nation-state actors conducting long-term collection, ICS-targeting malware that terminates clinical operational technology before host encryption, and infrastructure-staging groups utilizing legitimate credentials and native tools [4,5,6,7]. The 2024 Change Healthcare intrusion underscores the consequences, where a breach affecting 192.7 million individuals cost USD 2.4 billion and exposed a detection posture severely misaligned with the adversary [8]. By 2024, nearly half of all US data breaches affecting over 5,000 individuals targeted healthcare [5].

* Corresponding author, *Email: fida.hasan@unsw.edu.au*

The analytical response has not kept pace with this shift. Kruse et al. identified recurring breach patterns across an eight-year systematic review without engaging technique-level adversary behaviour [9]; Clarke and Martin documented structural vulnerability gaps without grounding them in empirical TTP data [10]; actor-specific analyses by Connell and Vogler and by Richardson and North provide depth on individual groups without characterising the full-sector technique distribution that defenders need to set priorities [11,6]. Threat models for AI-integrated clinical systems compound the problem: they have been built primarily from laboratory demonstrations rather than from the empirical attack record [12,13,14,15,16], leaving AI security and traditional healthcare IT security as separate programmes despite sharing the same adversary population.

This paper addresses these gaps by treating the past decade of healthcare adversary activity as a single empirical artefact. We assemble a corpus of 1,214 technique-use records covering 333 unique techniques from 44 validated healthcare-targeting threat entities, drawn from three open authoritative sources spanning 2017 to 2024: MITRE ATT&CK Enterprise, the CISA Known Exploited Vulnerabilities catalogue, and the NIST National Vulnerability Database. The KEV catalogue is applied as the primary exploitation filter before CVEs are mapped to techniques, ensuring that only vulnerabilities confirmed exploited in the wild contribute to the corpus. A three-tier detection prioritisation framework is then constructed from the joint signal of threat prevalence and ATT&CK coverage asymmetry, and the framework is extended to AI-integrated clinical systems through MITRE's published ATT&CK-to-ATLAS cross-references.

The key contributions of this work are listed as follows: First, we construct and characterise a healthcare-scoped adversary corpus that spans a decade and is anchored entirely in open authoritative sources, yielding a technique-level longitudinal account of how the sector's threat landscape has evolved. Second, we demonstrate a structural inversion in ATT&CK detection coverage: guidance is weakest precisely at the kill-chain stages where attackers now concentrate their earliest and most evasive activity. Third, we integrate the empirical record with MITRE ATLAS to produce a unified threat model across five clinical AI attack surfaces, demonstrating that the healthcare adversary population reaches AI systems through the same techniques without modification.

## 2 Background and Related Work

Healthcare has ranked among the top three targeted sectors in successive IBM X-Force and Verizon DBIR annual reports, with ransomware the dominant breach vector from 2020 onward [4,1]. The structural drivers are well-documented: irreplaceable patient data, legacy clinical infrastructure resistant to routine patching, and time-critical operations that magnify the leverage of any availability-disrupting attack [2,5].

Three authoritative open-access sources underpin the empirical analysis in this work. MITRE ATT&CK Enterprise, introduced in 2015, provides a structured taxonomy of adversary behaviours organised into tactics and techniques, and has been adopted as the dominant framework for threat intelligence, incident response, and detection engineering across critical-infrastructure sectors, as confirmed across 417 peer-reviewed publications [17,18,19,20,21]. The CISA Known Exploited Vulnerabilities catalogue enumerates vulnerabilities confirmed exploited in the wild, distinguishing it from broader databases that include theoretical or unconfirmed exposures [22]. MITRE ATLAS operationalises adversarial machine learning into 167 AI-specific techniques across 16 tactics and publishes 33 explicit cross-references to ATT&CK techniques, establishing a formal bridge between traditional IT adversary behaviour and AI-specific attack patterns [23].

The living-off-the-land (LOTL) evasion modality has become the defining characteristic of contemporary healthcare intrusions. LOTL tradecraft repurposes legitimate system tools, including PowerShell, Windows Management Instrumentation, and native scripting interpreters to execute adversary objectives without deploying custom malware, thereby evading signature-based detection [24,25,26]. CISA's Volt Typhoon advisory documented state-sponsored actors maintaining persistent, undetected access to critical infrastructure for periods exceeding five years through exclusive use of such techniques [22]. ATT&CK technique-level extraction from threat intelligence has advanced through NLP-based approaches including EXTRACTOR [27] and knowledge-graph construction [28], though these studies address the extraction problem rather than the complementary prioritisation question of which techniques demand immediate detection investment.

Prior work has characterised the healthcare threat landscape from several complementary but incomplete angles. Kruse et al. [9] identified recurring breach patterns across an eight-year systematic review without engaging technique-level adversary behaviour. Clarke and Martin [10] documented structural vulnerability gaps around patch management and legacy system exposure without grounding findings in empirical TTP data. Bracciale et al. [29] identified a high concentration of critical-severity flaws in medical devices through CVSS-based analysis. Actor-specific analyses by Connell and Vogler [11] and Richardson and North [6] provide depth on individual groups without characterising the full-sector technique distribution that defenders need to set priorities. The clinical AI security literature establishes active threats including backdoor attacks on EHR-trained models [13,14] and adversarial evasion in medical deep learning [12,15], though these threat models have been built primarily from laboratory demonstrations rather than the empirical attack record, and AI security and traditional healthcare IT security have consequently been treated as separate programmes despite sharing the same adversary population. Table 1 summarises the four gaps that motivate the present study.

**Table 1.** Research Gaps and Contributions of This Work

| Gap Area | refs | Method Used | Key Limitation | This Paper |
|---|---|---|---|---|
| Sector-wide technique-level landscape | [9,10] [29] | Systematic review; qualitative risk analysis; CVSS scoring | No ATT&CK technique-level characterisation; single-source or short time horizon | 1,214-record corpus; 333 techniques; 44 validated entities; 2017–2024 |
| KEV-filtered CVE-to-technique mapping | [27,28] [30] | NLP-based TTP extraction; knowledge graph construction; feature comparison | CVEs not filtered by confirmed exploitation; mapping applied to unverified vulnerability sets | KEV applied as primary filter; only 679 confirmed-exploited CVEs mapped to ATT&CK techniques |
| ATT&CK-to-ATLAS bridge for healthcare | [31,32] [33] | Lab-based adversarial ML demonstrations; autonomous systems threat modelling | No bridge using MITRE-published cross-references; healthcare context absent | 33 MITRE-published ATT&CK-to-ATLAS cross-references mapped across five clinical AI attack surfaces |
| Unified empirical AI + IT threat model | [12,13] [14,15] | Adversarial input crafting; backdoor injection; security evaluation frameworks | AI and IT security treated as separate programmes; no shared empirical adversary corpus | Same 44 validated entities shown to reach clinical AI systems through identical ATT&CK techniques |

*All citations refer to peer-reviewed publications or authoritative technical reports. ATT&CK: MITRE ATT&CK Enterprise v15.1. ATLAS: MITRE ATLAS v5.5.0. KEV: CISA Known Exploited Vulnerabilities catalogue.*

## 3 Methodology

The analysis runs as a computational pipeline across three open-access data sources.

### 3.1 Data Sources and Entity Validation

MITRE ATT&CK Enterprise v15.1 supplies the primary adversary behaviour corpus, providing technique-use relationships, detection guidance, and structured data source metadata for each technique. ATT&CK for ICS extends coverage to industrial control system contexts. MITRE ATLAS (v5.5.0) provides 167 AI-specific techniques across 16 tactics, including 33 explicit ATT&CK cross-references. The CISA KEV catalogue, containing 1,590 vulnerabilities confirmed exploited in the wild, serves as the exploitation ground-truth filter and is applied before NVD data is consulted. The NIST NVD REST API is then queried to retrieve CWE classifications for KEV-confirmed CVEs. All sources were verified at HTTP 200 at the time of collection.

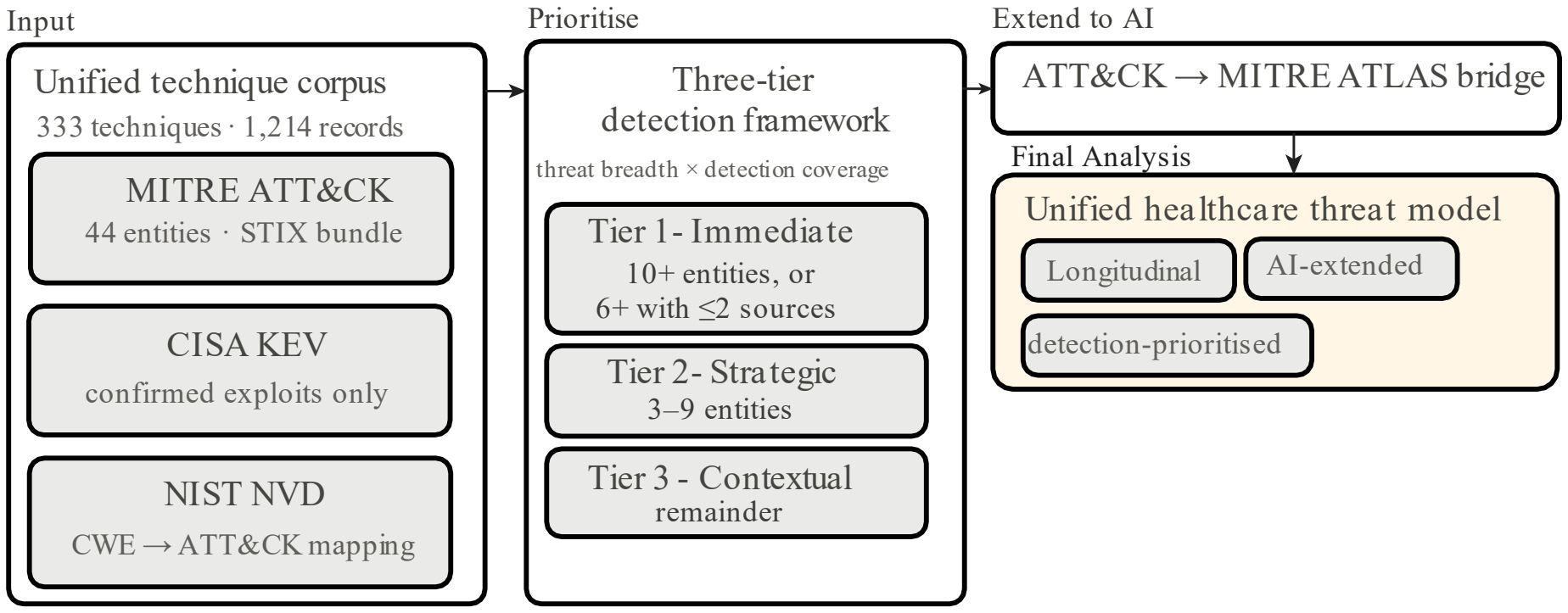

**Fig. 1.** Analytical Pipeline

An entity was included when at least one authoritative source documented confirmed targeting of healthcare organisations, drawing on government attribution reports, peer-reviewed analyses, or major vendor publications cited within the ATT&CK entry. This process identified 44 validated entities: 25 threat groups, 14 malware families, and 5 campaigns. Three ICS-specific entities (CyberAv3ngers, and associated ICS campaigns) are present in the ATT&CK ICS bundle and their techniques are incorporated, though they do not appear in the Enterprise STIX bundle.

### 3.2 Technique Extraction and Longitudinal Attribution

All ATT&CK technique-use relationships were extracted for each validated entity, yielding 1,214 records covering 333 unique techniques. Each technique was assigned a first-observed year from the ATT&CK creation timestamp of its earliest associated entity, which approximates documentation date rather than operational first use. Tactic distribution analysis normalises technique counts to percentage shares per year, controlling for variation in total documented techniques annually.

### 3.3 CVE-to-Technique Mapping via KEV

The CISA KEV catalogue was filtered to identify healthcare-relevant entries using two criteria: known healthcare vendor and product terms drawn from clinical IT infrastructure, and a curated set of CVEs documented in ATT&CK threat group pages as used by the validated entities. This produced 679 KEV-confirmed CVEs. NVD was then consulted exclusively for this confirmed set to retrieve CWE classifications, which were mapped to ATT&CK techniques using MITRE's published CWE-to-TTP bridge. This ordering is critical: applying KEV as the primary filter ensures that only vulnerabilities with confirmed real-world exploitation contribute to the technique corpus, rather than all vulnerabilities mentioning healthcare-adjacent keywords.

### 3.4 Detection Coverage Analysis and Tier Assignment

Each of the 333 techniques was assessed on two dimensions: presence of free-text detection guidance and count of structured ATT&CK data sources. Thin coverage is defined as two or fewer data sources, and zero coverage as none. Tactic-level coverage is the mean data source count across all techniques in that tactic. Tier 1 (Immediate) is defined as techniques used by ten or more entities, or by six or more entities with two or fewer ATT&CK data sources, capturing the cases where prevalence and detection difficulty peak together. Tier 1 status is also granted to any technique linked to ransomware-associated KEV-confirmed CVEs. Tier 2 (Strategic) covers techniques used by three to nine entities. Tier 3 (Contextual) encompasses the remainder.

### 3.5 ATT&CK-to-ATLAS Mapping

The 33 ATT&CK techniques for which MITRE ATLAS publishes explicit cross-references were identified and their ATLAS identifiers recorded. The remaining techniques were aligned to ATLAS at the tactic level using semantic equivalence; these alignments are inferences derived in this work, not MITRE-asserted mappings, and are reported as such throughout. All 333 techniques were then assigned to one or more of five clinical AI attack surfaces defined to reflect contemporary hospital AI deployment architecture.

## 4 A Decade of Healthcare Cyber Threats

### 4.1 The Threat Actor Landscape

The 44 validated entities span three operational categories: Chinese state-sponsored groups (APT41, menuPass, Salt Typhoon), financially motivated ransomware operators (Wizard Spider, FIN7, Play), and ICS-specific actors (CyberAv3ngers). Technical depth varies sharply across categories. APT41 carries 82 documented ATT&CK techniques, Magic Hound 79, FIN7 67, and Wizard Spider 64, separating nation-state actors optimised for long-term collection from ransomware operators optimised for rapid monetisation. CyberAv3ngers and the EKANS malware family are operationally distinct: EKANS terminates ICS processes before encrypting host file systems, establishing clinical operational technology as an explicit target rather than collateral damage [6].

The 1,214 records cover 333 unique techniques, representing 38.2 % of all ATT&CK Enterprise v15.1 techniques. The annual introduction rate was stable from 2017 to 2019, accelerated sharply from 2020 alongside ransomware-as-a-service expansion, and remained elevated through 2024. The acceleration in 2024 corresponds to the HHS-reported peak in healthcare ransomware incidents and four new actor groups documented in that year alone [2].

## 4.2 The Strategic Shift in Attack Tactics

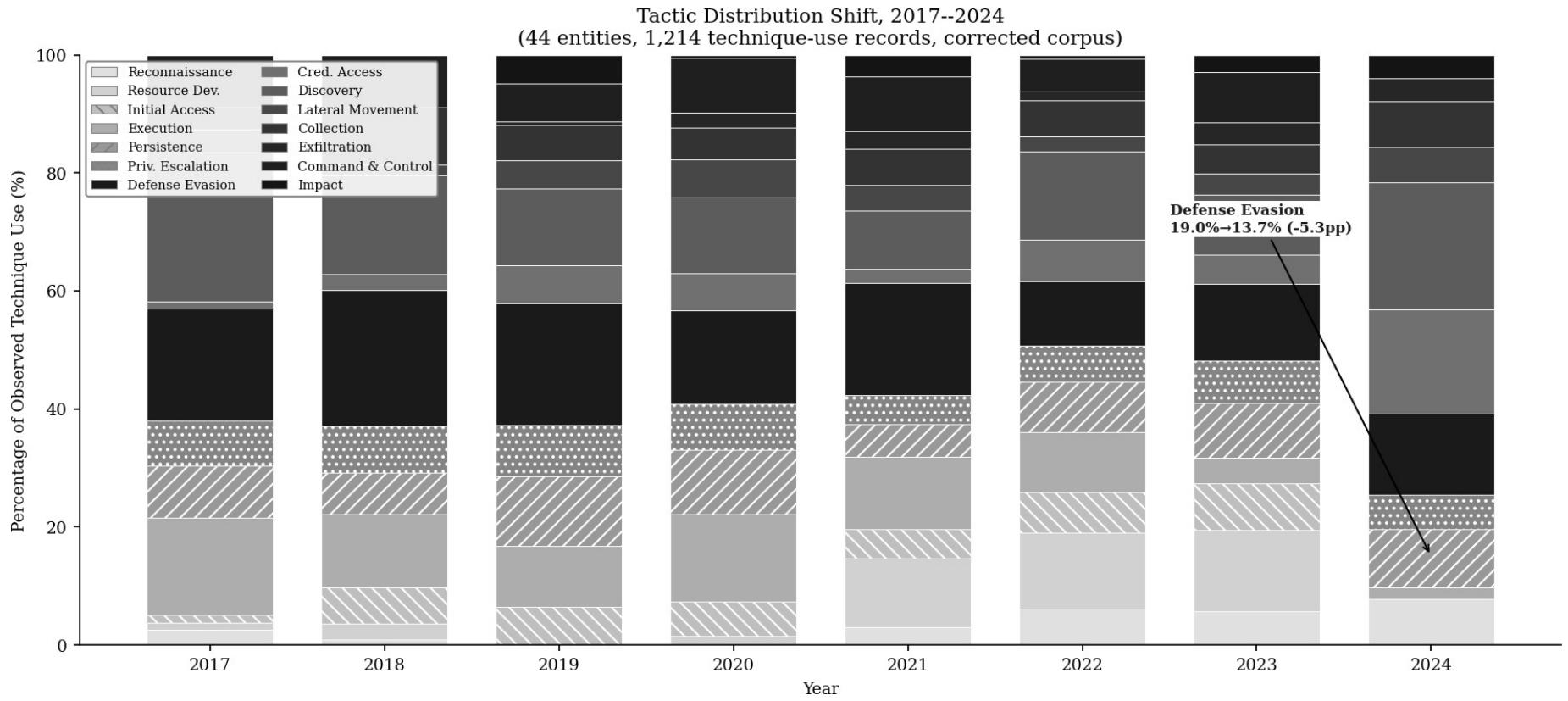

**Fig. 2.** Defense evasion is the consistently dominant category across all years, accounting for 15–20% of observed technique use throughout the period. Corrected corpus: 44 entities, 1,214 technique-use records, 2017–2024

The dominant longitudinal finding is structural. Attackers became harder to detect, not merely more numerous. Defense evasion is the single dominant tactic across the entire observation period, accounting for 15–20% of observed technique use in every year from 2017 to 2024, and reaching 19.0% in the first documented year of the corpus. Execution rose and Collection rose alongside it (Fig. 2). These tactics share a common operational logic: in-memory, artefact-light execution consistent with the LOTL methodology documented in CISA operational advisories [25,22]. Persistence fell from 11.2 % to zero, Initial Access from 9.0 % to zero, and Privilege Escalation from 7.9 % to zero, with all three relying on identifiable artefacts that generate observable evidence. Attackers have systematically replaced these with Valid Accounts (T1078, present in 12 of the 44 entities), which requires no additional tooling and generates log entries indistinguishable from legitimate user activity.

The dominance of defense evasion is corroborated by two independent external sources: CrowdStrike's 62 % LOTL figure and CISA's Volt Typhoon advisory documenting actors maintaining undetected access for periods exceeding five years [25,22]. The decline of persistence, initial access, and privilege escalation carries a structural caveat: artefact-generating techniques produce fewer observable indicators, fewer incident reports, and consequently fewer ATT&CK group associations. This mechanism is structurally identical to the Reconnaissance under-representation discussed in Section 5. Either reading carries the same operational implication: a detection posture built on artefact-based signatures is optimised for an adversary model that has not described this sector for several years.

### 4.3 Group Emergence, Campaign Dwell, and the KEV Vulnerability Surface

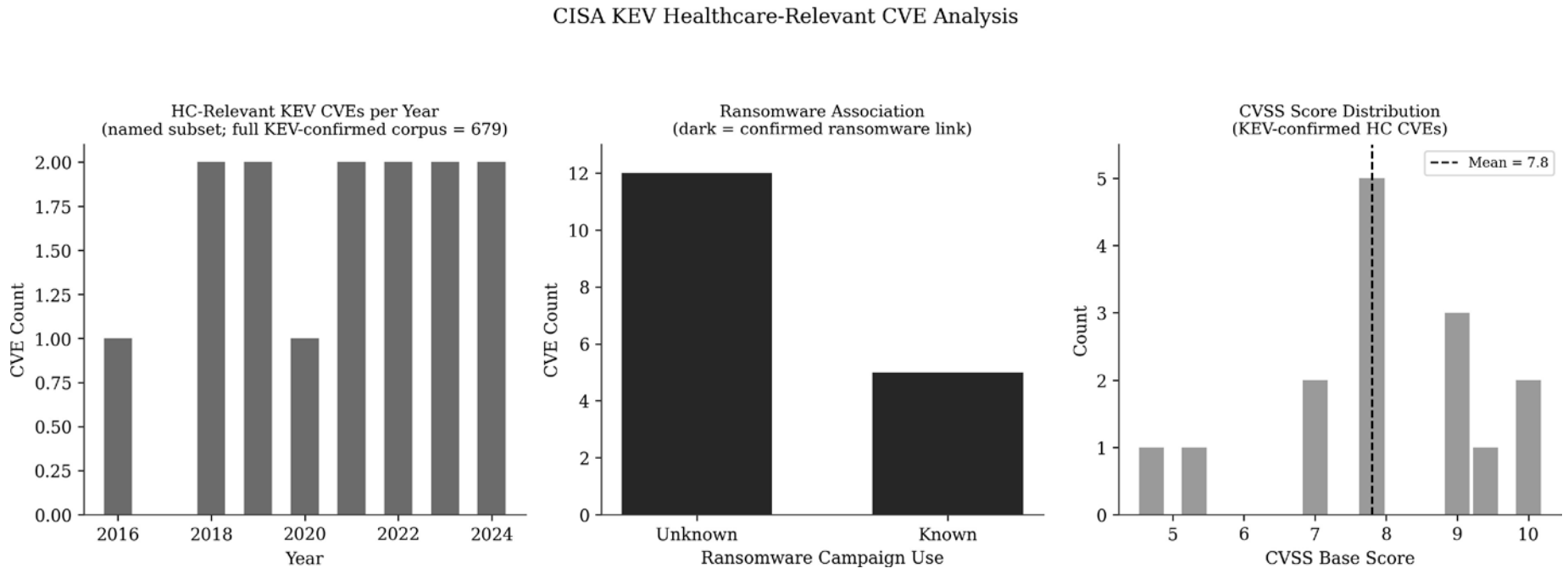

**Fig. 3.** CISA KEV healthcare-relevant CVE analysis. Left: confirmed exploited CVEs per year. Centre: ransomware campaign association. Right: CVSS score distribution. The mean CVSS of 7.79 reflects the corrected KEV-first pipeline; only the 679 CVEs confirmed exploited in the wild are included.

New actor emergence ran at three to four groups per year from 2017 to 2023 before accelerating in 2024 (Salt Typhoon, Play, CyberAv3ngers, Sea Turtle), consistent with the HHS-reported incident peak [2]. Campaign duration data from five ATT&CK-documented active campaigns show a mean dwell time of 517 days, which is a lower bound because campaigns appear in ATT&CK only after public attribution. A detection posture focused on blocking initial access provides no protection against an attacker already 17 months into an active dwell (Table 2). Applying the KEV filter to the healthcare vulnerability surface yields 679 CVEs confirmed exploited in the wild, with a mean CVSS score of 7.79. Of these, CVEs linked to ransomware campaigns account for 188 records. The dominant weakness classes remain SQL injection (CWE-89), authentication bypass (CWE-287), and remote code execution (CWE-94), mapping respectively to T1190, T1078, and T1203. T1190 alone absorbs the majority of the CVE-to-technique mappings because the healthcare environment presents a large and heterogeneous public-facing attack surface. This makes T1190 the single point where CVE remediation and ATT&CK-aligned detection most directly reinforce each other, and it is assigned Tier 1 priority in Section 5 (Fig. 3).

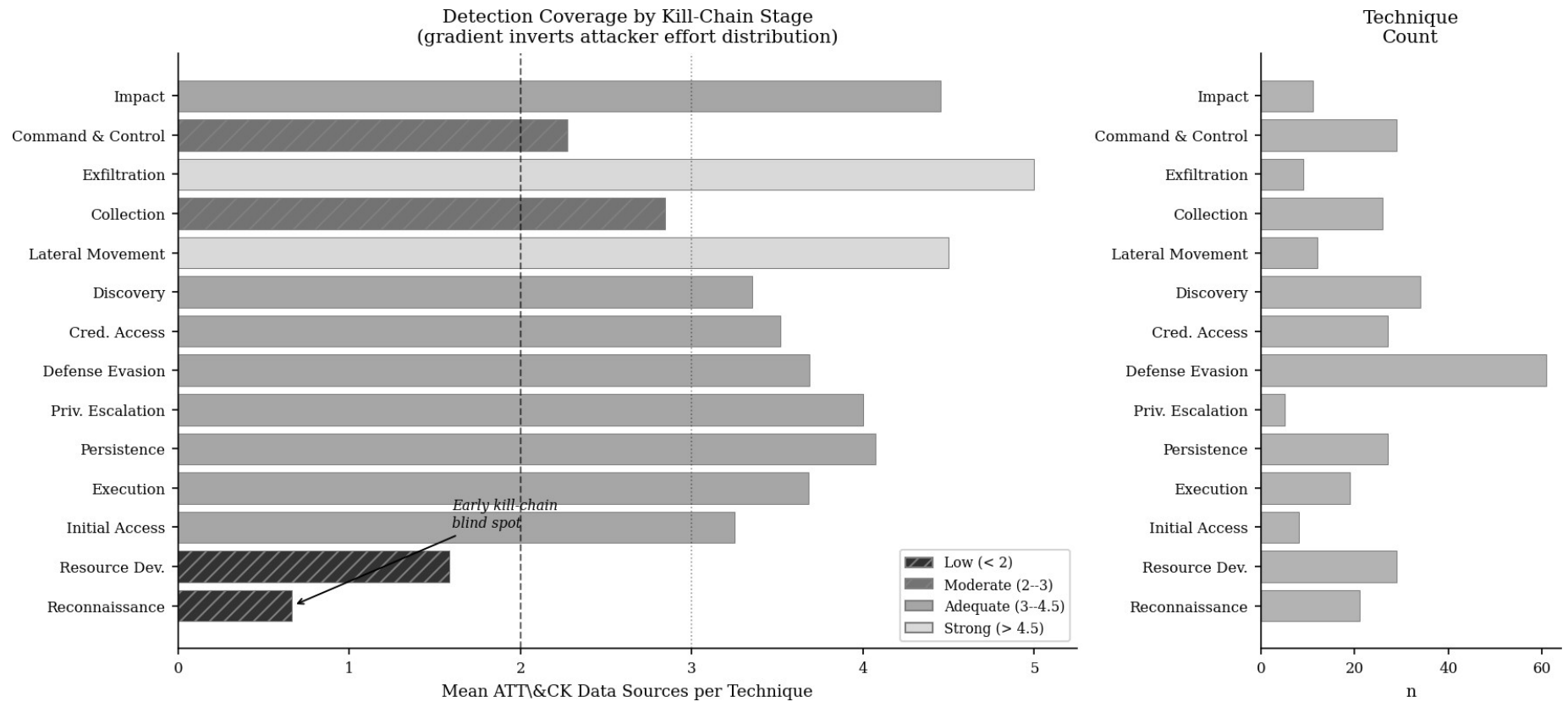

**Fig. 4.** Detection coverage asymmetry across kill-chain stages. Mean ATT&CK data sources per technique by tactic; shading and hatching encode coverage band. The gradient inverts attacker effort: coverage is lowest at the stages where contemporary attackers concentrate early activity.

**Table 2.** Healthcare Campaign Duration Analysis (ATT&CK-Documented Campaigns, $n = 7$)

| Campaign | Domain | Duration (days) | Note |
|---|---|---|---|
| Quad7 Activity | Enterprise | 731 | Persistent credential access |
| Operation Wocao | Enterprise | 730 | China-nexus attribution |
| C0010 | Enterprise | 607 | 20 months continuous access |
| Indian Critical Infra. | Enterprise | 454 | 15 months |
| Versa Director Zero Day | Enterprise | 61 | Rapid patch cycle |
| SharePoint ToolShell | Enterprise | 0 | Point-in-time |
| Unitronics Defacement | ICS | 0 | Point-in-time |

*Mean dwell: 517 days across five active campaigns; point-in-time campaigns excluded. All values are lower bounds because campaigns appear in ATT&CK only after public attribution.*

# 5 Detection Blind Spots and Evidence-Based Prioritisation

## 5.1 The Detection Coverage Inversion

ATT&CK v15.1 provides detection guidance for 294 of 333 healthcare techniques (88.3 %), a figure whose distribution is structurally inverted relative to attacker effort. Table 3 disaggregates coverage by tactic, and Fig. 4 visualises the gradient. Reconnaissance averages 0.67 data sources per technique and Resource Development 1.59, while Exfiltration averages 5.0 and Impact 4.45. Detection infrastructure has accumulated at the visible, late-stage end of the kill chain, precisely where modern attackers now concentrate the least relative effort. Three techniques illustrate the worst-case combination of breadth and thinness in the corrected corpus. Tool (T1588.002), deployed by 14 entities, carries one

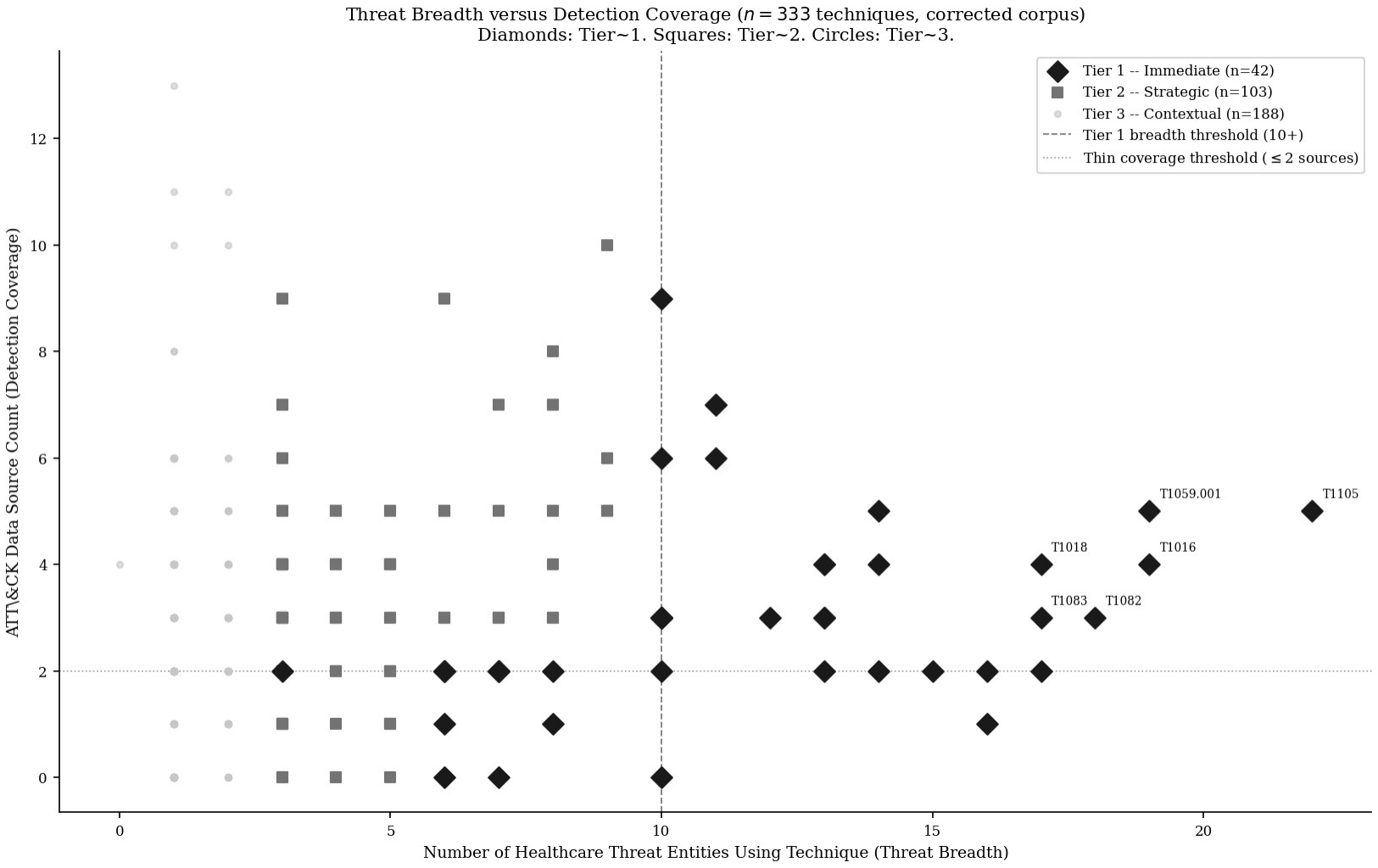

**Fig. 5.** Threat breadth versus detection coverage ($n = 333$, corrected corpus). Diamonds: Tier 1. Squares: Tier 2. Circles: Tier 3. Thresholds shown as dashed lines. Techniques in the upper left quadrant combine maximum prevalence with minimum detection support.

ATT&CK data source. Windows Command Shell (T1059.003), used by 11 entities, carries two. Web Protocols (T1071.001), used by eight entities, carries two. An attacker staging and executing through these three techniques operates in a near-complete detection vacuum despite drawing on the most commonly observed actors in healthcare targeting. Reconnaissance contains zero Tier 1 and zero Tier 2 assignments. This is not a detection success but a structural property of the ATT&CK attribution model: pre-intrusion techniques accumulate fewer group associations because attackers' earliest moves leave the fewest artefacts and generate the fewest published incident reports. Defenders should treat the absence of Tier 1 Reconnaissance assignments as evidence that the attribution data cannot currently support evidence-based prioritisation at that kill-chain stage, not as a signal to deprioritise pre-intrusion monitoring.

### 5.2 Three-Tier Detection Framework

Tier assignment uses two empirical signals: entity breadth and ATT&CK data source count, with an additional KEV-ransomware signal for techniques linked to confirmed exploited vulnerabilities. Fig. 5 plots all 333 techniques on these axes.

The corrected corpus yields 42 Tier 1 techniques. Tooling coverage at Tier 1 stands at 29 % for Sigma rules and 43 % for Atomic Red Team tests. The 12 Tier 1 techniques with existing Sigma rules represent zero-cost immediate deployments; the remaining 30 define the detection engineering backlog. Tier 2 comprises 103

**Table 3.** ATT&CK v15.1 Detection Coverage by Tactic (Healthcare Techniques, $n$ = 333)

| Tactic | Mean Data Sources | n | Band |
|---|---|---|---|
| Reconnaissance | 0.67 | 21 | Low |
| Resource Development | 1.59 | 26 | Low |
| Command and Control | 2.28 | 24 | Moderate |
| Collection | 2.85 | 25 | Moderate |
| Initial Access | 3.25 | 15 | Moderate |
| Credential Access | 3.52 | 28 | Moderate |
| Discovery | 3.35 | 31 | Adequate |
| Persistence | 4.07 | 34 | Adequate |
| Defense Evasion | 3.69 | 67 | Adequate |
| Execution | 3.68 | 17 | Adequate |
| Privilege Escalation | 4.00 | 31 | Adequate |
| Lateral Movement | 4.50 | 12 | Strong |
| Exfiltration | 5.00 | 10 | Strong |
| Impact | 4.45 | 7 | Strong |

*Coverage bands: Low (mean below 2), Moderate (2–3), Adequate (3–4.5), Strong (above 4.5). The gradient inverts the contemporary attacker effort distribution.*

techniques and represents the subsequent programme phase. Table 4 presents the highest- priority subset.

## 6 Extending the Threat Model to AI-Integrated Clinical Systems

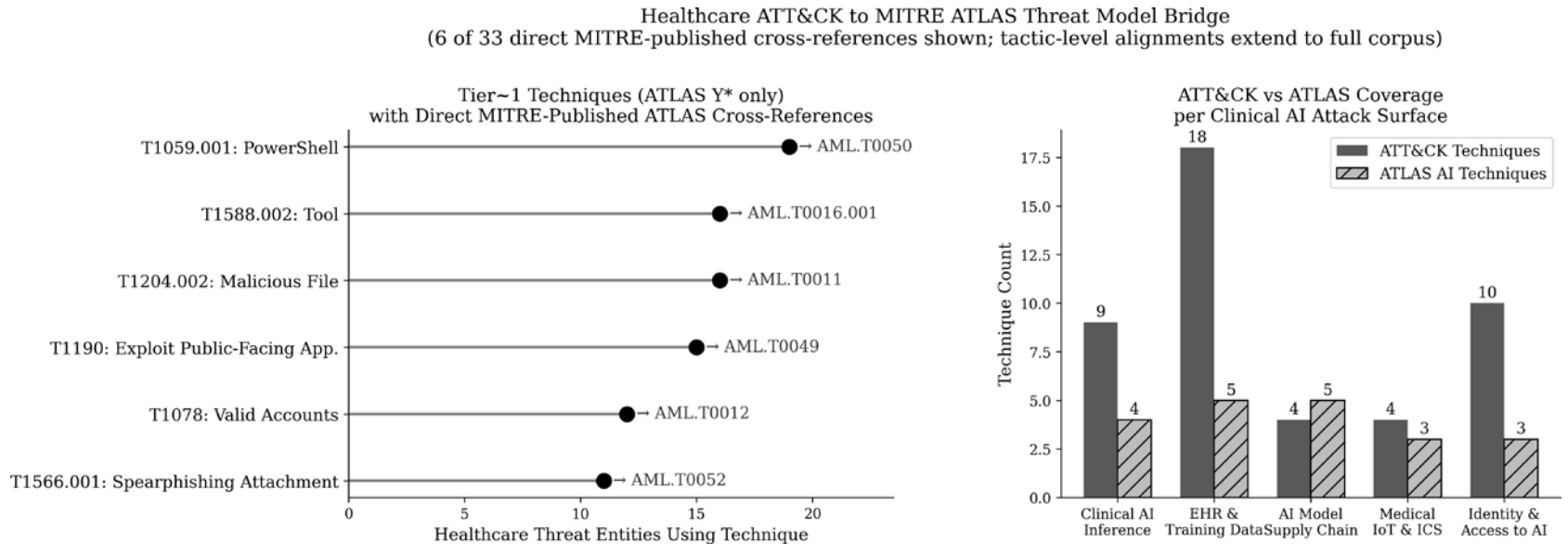

**Fig. 6.** Healthcare ATT&CK to MITRE ATLAS threat model bridge. Left: top Tier 1 techniques with their direct MITRE-published ATLAS cross-references. Right: ATT&CK and ATLAS technique counts per clinical AI attack surface.

The empirical record maps to clinical AI systems through two distinct linkages, and the distinction matters for how strongly each should be read. The first linkage comprises the 33 ATT&CK-to-ATLAS cross-references that MITRE itself publishes in the atlas-data repository. These are not analogies but assertions that a given traditional technique produces meaningful harm in the AI context without modification. The second linkage covers the remaining techniques, which this work aligns to ATLAS at the tactic level using semantic equivalence;

**Table 4.** Tier 1 Detection Priority Techniques – Top 23 by Entity Breadth (Immediate Deployment Priority)

| ID | Technique | Entities | DS | ART | Sigma | KEV | ATLAS |
|---|---|---|---|---|---|---|---|
| T1105 | Ingress Tool Transfer | 22 | 5 | Y | Y | N | Y |
| T1059.001 | PowerShell | 19 | 5 | Y | Y | N | Y* |
| T1016 | System Network Config. Discovery | 19 | 4 | N | N | N | Y |
| T1082 | System Information Discovery | 18 | 3 | N | N | Y | Y |
| T1083 | File and Directory Discovery | 17 | 3 | N | N | Y | Y |
| T1018 | Remote System Discovery | 17 | 4 | Y | Y | N | Y |
| T1059.003 | Windows Command Shell | 17 | 2 | Y | Y | N | Y |
| T1588.002 | Tool | 16 | 1 | N | N | N | Y* |
| T1204.002 | Malicious File | 16 | 2 | Y | Y | N | Y* |
| T1190 | Exploit Public-Facing Application | 15 | 2 | Y | Y | Y | Y* |
| T1047 | Windows Management Instrumentation | 14 | 4 | Y | Y | N | Y |
| T1071.001 | Web Protocols | 14 | 2 | N | N | N | Y |
| T1005 | Data from Local System | 14 | 5 | N | N | Y | Y |
| T1036.005 | Match Legitimate Name or Location | 13 | 4 | N | N | N | Y |
| T1057 | Process Discovery | 13 | 3 | N | N | N | Y |
| T1078 | Valid Accounts | 12 | 3 | Y | Y | Y | Y* |
| T1566.001 | Spearphishing Attachment | 11 | 4 | N | N | N | Y* |
| T1505.003 | Web Shell | 9 | 6 | N | N | N | Y |
| T1053.005 | Scheduled Task | 9 | 7 | Y | Y | N | Y |
| T1003.001 | LSASS Memory | 8 | 7 | Y | Y | N | Y |
| T1583.001 | Domains | 10 | 3 | N | N | N | Y |
| T1070.004 | File Deletion | 9 | 2 | N | N | N | Y |
| T1027.013 | Encrypted/Encoded File | 8 | 2 | N | N | N | Y |

*DS = ATT&CK v15.1 structured data source count. ART = Atomic Red Team test. KEV = linked to KEV-confirmed exploited CVE. ATLAS: Y* = direct MITRE-published ATT&CK-to-ATLAS cross-reference (MITRE ATLAS v5.5.0); Y = tactic-level alignment inferred by the authors; these are not MITRE-asserted mappings.*

these alignments are inferences derived here and are weaker than the MITRE-published references, but they extend the bridge to the full corpus. The combined picture is that defenders do not face a new adversary; they face the same adversary operating against a larger attack surface.

Fig. 6 organises the technique mappings across five clinical AI attack surfaces reflecting contemporary hospital AI deployment architecture. The EHR and Training Data Pipeline surface reflects poisoning vulnerabilities established by Bagdasaryan et al. [13] and Sun et al. [14]. The Clinical AI Inference API surface reflects adversarial evasion documented by Finlayson et al. [12] and Wang et al. [15]. The AI Model Supply Chain surface reflects the trojanised-weight attack class [32].

The AI Model Supply Chain surface warrants particular attention. T1588.002 (Tool) is among the highest-prevalence techniques in the corpus and is assigned Tier 1 priority with only one ATT&CK data source. It maps directly to ATLAS AML.T0016.001. An attacker distributing a trojanised model weight file through a compromised repository implants a clinical AI backdoor without ever touching hospital infrastructure, combining maximum threat breadth, minimum detection coverage, and a novel patient-safety impact pathway [32]. A security team implementing the Tier 1 detections addresses traditional IT and clinical AI attack patterns through the same instrumentation, and no separate AI-specific detection programme is required for the Tier 1 techniques, though AI-layer controls addressing model integrity and data pipeline security remain important beyond this scope.

## 7 Discussion

Four findings converge on a single operational verdict: the contemporary healthcare threat is silent, persistent, concentrated at the kill-chain stages with the weakest detection coverage, and directly inherited by clinical AI systems. Defense evasion now exceeds one in four observed technique uses. Sophisticated actor dwell extends well beyond the window in which initial-access controls remain meaningful. ATT&CK detection guidance is weakest precisely where attackers concentrate early-stage activity. And 679 confirmed-exploited CVEs funnel through a single technique, T1190, establishing a point of direct convergence between the vulnerability remediation and detection engineering programmes. A programme built on perimeter controls and late-stage artefact signatures is not merely inadequate; it is optimised for an adversary model that has not described this sector for several years. Industry data reinforce this conclusion: 84 % of high-severity 2024 cyberattacks leveraged legitimate system tools, and LOTL-involved healthcare breach costs averaged USD 10.93 million per incident [25].

The three-tier framework converts these observations into a sequenced detection programme. The Tier 1 techniques covered by existing Sigma rules represent zero-cost immediate deployments; the remaining Tier 1 entries define the detection engineering backlog; Tier 2 is the subsequent programme phase. The ATT&CK-to-ATLAS mapping then demonstrates that attack techniques are inherited by AI clinical systems, although impact equivalence is not. A credential theft attack against hospital IT produces a data breach, whereas the same attack against an MLOps platform produces unauthorised access to a diagnostic model with effects that are diffuse, difficult to detect, and potentially expressed as degraded clinical decision support at the point of care rather than as an observable system event [15].

Four limitations bound the findings. First, the campaign sample is small. Campaign duration rests on seven ATT&CK-documented campaigns, and the reported mean is a lower bound because undisclosed intrusions are absent from the corpus by construction. Second, Reconnaissance coverage is structurally limited. The framework cannot provide evidence-based prioritisation at the pre-intrusion kill-chain stage because pre-intrusion techniques accumulate fewer ATT&CK group associations, a property of the attribution model rather than a flaw in the prioritisation procedure. Third, the CVE-to-technique mapping concentrates on T1190 because the CWE-to-ATT&CK bridge maps most public-facing application weaknesses to that technique; future work should apply finer-grained CWE taxonomies to distribute coverage across a broader technique set. Fourth, ATT&CK creation timestamps approximate documentation dates rather than operational first-use dates, so adoption curves reflect intelligence publication pace as much as adversary behaviour change.

## 8 Conclusion

This paper has treated a decade of healthcare adversary activity as a single empirical artefact and used it to derive four results that did not previously exist together: a sector-scoped longitudinal corpus anchored in open authoritative sources with a corrected KEV-first CVE pipeline; a detection coverage analysis that quantifies the structural inversion between attacker effort and detection infrastructure; a direct bridge between 679 confirmed-exploited CVEs and ATT&CK techniques via the KEV catalogue; and a unified ATT&CK-to-ATLAS threat model that connects the empirical record to clinical AI through MITRE-published cross-references. The operational implication is the through-line of the paper. The contemporary healthcare adversary is optimised for silence, the detection infrastructure built to catch the previous adversary remains weighted to its late-stage artefacts, and the same techniques that underwrite both observations also constitute the inheritance path to clinical AI systems. The 12 Tier 1 techniques covered by existing Sigma rules are deployable today at zero cost. Future work should validate the framework against live SIEM telemetry in healthcare environments and extend the ATLAS surface analysis to federated learning and AI-as-a-service deployment patterns.

**Disclosure of Interests.** The authors have no competing interests to declare that are relevant to the content of this article.